\documentclass[pra,showpacs,showkeys,superscriptaddress,amssymb,amsmath]{revtex4}
\usepackage{graphicx}
\usepackage{hyperref}
\usepackage{placeins}
\usepackage{xcolor}
\usepackage{caption}
\usepackage{listings}
\usepackage[normalem]{ulem}
\usepackage{color}
\usepackage{float}
\usepackage{simplewick}

\begin{document}

	\title{An exact solution method for the enumeration of connected Feynman diagrams  
	}
	
	\author{E. R. Castro}
	\email[]{ercmlv@gmail.com}
	\noaffiliation

	\author{I. Roditi}
	\email[]{roditi@cbpf.br}
	\affiliation{Centro Brasileiro de Pesquisas F\'{\i}sicas/MCTI,
		22290-180, Rio de Janeiro, RJ, Brazil}
	\affiliation{Institute for Theoretical Physics, ETH Zurich 8093, Switzerland}

	\begin{abstract}
		We completely generalize previous results related to the counting of connected Feynman diagrams. We use a generating function approach, which encodes the Wick contraction combinatorics of the respective connected diagrams. Exact solutions are found for an arbitrary number of external legs, and a general algorithm is implemented for this calculus. From these solutions, we calculate many asymptotics expansion terms for a simple analytical tool (Taylor expansion theorem). Our approach offers new perspectives in the realm of Feynman diagrams enumeration (or zero-dimensional quantum field theory).        
	\end{abstract}	
	\maketitle

	\section{Introduction} \label{Int}
	
The generating function approach is a standard way in which many problems of mathematics and physics are formulated \cite{Flajolet}. An exact solution of a particular problem can be obtained by powerful methods that, generally, result in a closed expression of this generating function. For example, in equilibrium statistical mechanics, the generating function is the partition function, from which, all the other relevant quantities can be calculated \cite{Baxter}. Non-closed expressions for the generating function, generally, admit only approximate methods or, in the best case, an algorithmic way to calculate the associated results. 

In zero dimensions, the generating functional of the correlation functions in QFT, which is formulated in terms of the classical action, converts into the generating function of the number of Feynman diagrams. That is a well know result \cite{Cvitanovic} and the powerful machinery of QFT calculus can be used to obtain a diversity of ways to count Feynman diagrams (for example, using Dyson and many-body equations, see references\cite{Argyres}, \cite{Molinari} and \cite{Pavlyukh}). In this respect, zero-dimensional field theory is understood as a toy model of realistic finite dimension theories where precise calculations can be realized. In a more mathematical context, the algebraic theory of formal power series also has been used, showing explicitly the rich combinatorial structure present in the Feynman diagrams and expressed in modern enumerative combinatorial language. Also, asymptotic mathematics finds here a fertile source of problems where powerful mathematics techniques can be tested: the generating function is generally expressed in function of exponential integrals, and complex variable tools can be applied. To learn more about these mathematical approaches, see \cite{Borinsky}, \cite{Abdesselam}, \cite{Jackson}, \cite{Yeats}, \cite{Geloun} and references therein.       

Recently (see refs \cite{Prunotto}, \cite{Gopala} and \cite{Castro}), the counting of connected Feynman diagrams in many-body theory (or in scalar electrodynamics) was related to the enumeration of rooted maps (combinatorial objects represented as graphs embedded into bidimensional compact orientable surfaces \cite{Walsh}). The important thing to stress is that there exist an explicit counting formula, obtained by ref.\cite{Arques}, for the enumeration of these objects. The counting formula for the $m$-order connected Feynman diagrams with two external legs and for the $m$-edges rooted maps is

\begin{equation}\label{conn}
\mathfrak{h}_{m}^{(1)}=\frac{1}{2^{m+1}}\sum_{i=0}^{m}(-1)^{i}\sum_{a_{1}=1}^{\infty}\cdots\sum_{a_{i+1}=1}^{\infty}\delta_{a_{1}+\cdots+a_{i+1},m+1}\prod_{j=1}^{i+1}\frac{(2a_{j})!}{a_{j}!}.
\end{equation}

This formula is an alternating finite sum whose terms are indexed by the different ways to sum (considering the order) the number $m+1$. In other words, the sum is expressed as a function of the compositions of the number $m+1$

\begin{equation}\label{conn2}
\mathfrak{h}_{m}^{(1)}=\frac{1}{2^{m+1}}\sum_{i=1}^{m+1}(-1)^{i+1}\sum_{c \in \mathfrak{C}_{i}(m+1)}F(c),
\end{equation}
where $c=(a_{1},a_{2},\cdots,a_{i})$ satisfied $a_{1}+a_{2}+\cdots+a_{i}=m+1$, and $\mathfrak{C}_{i}(m+1)$ is the set of compositions of $m+1$ with $i$ elements. The function $F(c)$ is simply

\begin{equation}\label{F}
F(c)=\prod_{j=1}^{i}\frac{(2a_{j})!}{a_{j}!}
\end{equation}

Many values of $\mathfrak{h}_{m}^{(1)}$ can be calculated with a computer. Unfortunately, for large $m$, the exact formula is completely useless since the number of compositions grows exponentially in $m$. A more efficient form to compute the values of $\mathfrak{h}_{m}^{(1)}$ is using a recursive approach introduced in ref \cite{Castro3}. In this reference, explicit recurrences were introduced in order to compute the number of connected Feynman diagrams with an arbitrary number of external legs. Those recurrences were solved only for the cases $N=1$ and $N=2$ (two and four external legs) leading to exact formulas expressed in the composition number form of formula (\ref{conn2}). The particular advantage of these formulas is that they allow a simple calculation of many terms of the asymptotic expansion. In the limit, $m\to \infty$ a new asymptotic contribution (negligible compared to the principal contribution) is explicitly calculated. In this paper, we generalize the results obtained in ref \cite{Castro3} to arbitrary $N$ (or for an arbitrary number of external legs). We also obtain an algorithm that allows us to calculate exact formulas for $N>2$ and as well as terms of their associated asymptotic expansions.

\section{The generating function of connected Feynman diagrams}

\begin{figure}[H]
	\centering
		\includegraphics[width=0.7\textwidth]{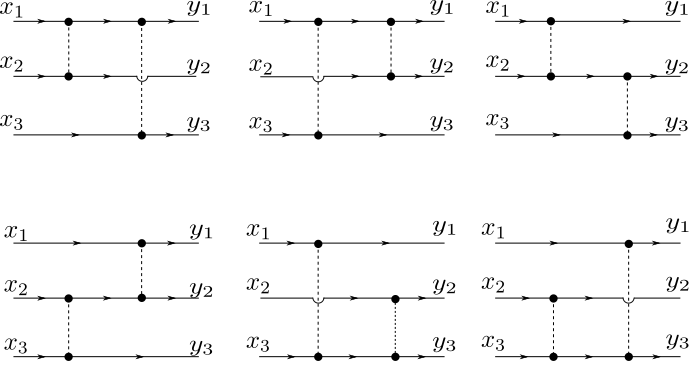}
	\caption{The six connected Feynman diagrams for $m=2$ and $N=3$. Note that the external legs are labeled. For unlabeled external legs, all these diagrams would be equivalent.}
	\label{Second6}
\end{figure}

In quantum field theory, the Wick theorem is the basis to construct the Feynman diagrams. Strictly speaking, the zero-dimension approach enumerate Wick contractions and the $m$-order Feynman diagrams are equivalence classes in the total set of $m$-order Wick contractions. We follow the conventions established in \cite{Castro3} for the counting of the many-body diagrams: $\mathcal{N}_{c\,m}^{(N)}$ correspond to the total number of $m$-order connected Wick contractions and $2N$ external legs. The number

\begin{equation}\label{h}
\mathfrak{h}_{m}^{(N)}=\frac{\mathcal{N}_{c\,m}^{(N)}}{2^{m}m!},
\end{equation}
counts $m$-order conected Feynman diagarms with labeled external legs. For $N=1$, $\mathfrak{h}_{m}^{(1)}$ counts explicitly unlabeled Feynman digrams. For $N>1$, the figure \ref{Second6} explains the label in the external legs.

The number of all possible $m$-order Wick contractions (connected and disconnected) with $2N$ external legs is given by \begin{equation}
\mathcal{N}_{m}^{(N)}=(2m+N)!.
\end{equation} 
The particular case $N=0$ is denoted as $\mathcal{D}_{m}=\mathcal{N}_{m}^{(0)}=(2m)!$ and enumerate all the vacuum Wick contractions.  

The bivariate generating function corresponding to the number of all the Wick contractions is

\begin{equation}
Z(x,y)=\sum_{N=0}^{\infty}\sum_{m=0}^{\infty}\frac{\mathcal{N}_{m}^{(N)}}{N!}\frac{x^{N}}{N!}\frac{y^{m}}{m!}=\sum_{m=0}^{\infty}\mathcal{D}_{m}\frac{y^{m}}{m!}+\sum_{N=1}^{\infty}\sum_{m=0}^{\infty}\frac{\mathcal{N}_{m}^{(N)}}{N!}\frac{x^{N}}{N!}\frac{y^{m}}{m!}.
\end{equation}
calling $g(y)$ as $$g(y)=\sum_{m=0}^{\infty}\mathcal{D}_{m}\frac{y^{m}}{m!}$$

we can rewrite $Z(x,y)$ as

\begin{equation}\label{Z}
Z(x,y)=g(y)\left(1+\sum_{N=1}^{\infty}\left\{g^{-1}(y)\times\left[\sum_{m=0}^{\infty}\frac{\mathcal{N}_{m}^{(N)}}{N!}\frac{y^{m}}{m!}\right]\right\}\frac{x^{N}}{N!}\right)
\end{equation}  

Considering $$g^{-1}(y)=\sum_{m=0}^{\infty}h_{m}y^{m},$$ the coefficients $h_{m}$ by simple formal power series multiplication must satisfy

\begin{equation}
\sum_{m=0}^{\infty}\left(\sum_{n=0}^{m}\frac{\mathcal{D}_{n}}{n!}h_{m-n}\right)y^{m}=1
\end{equation}

This leads to one set of infinite equations between the coefficients $\mathcal{D}_{a}$ and $h_{b}$, in particular 

\begin{align}\label{Dm}
\mathcal{D}_{0}h_{0}&=1 \\
\sum_{n=0}^{m}\frac{\mathcal{D}_{n}}{n!}h_{m-n}&=0 \,\,\,\,\, \mathrm{for}\,\,\, m>0 \,\,\, (m \in \mathbb{N}) 
\end{align}
By induction (see Appendix \ref{App1}), can be proved that $h_{m}$ is

\begin{equation}\label{hm}
h_{m}=\sum_{i=1}^{m}(-1)^{i}\sum_{c \in \mathfrak{C}_{i}(m)}F(c)
\end{equation}
where we use the notation below formula (\ref{F}). Now, we focus on equation (\ref{Z}). Here, it becomes evident a multiplication of two power series. Explicitly 

\begin{equation}\label{ghy}
g^{-1}(y)\times\left[\sum_{m=0}^{\infty}\frac{\mathcal{N}_{m}^{(N)}}{N!}\frac{y^{m}}{m!}\right]=\sum_{m=0}^{\infty}\left(\sum_{n=0}^{m}\frac{\mathcal{N}_{n}^{(N)}}{N!n!}h_{m-n}\right)y^{m}=G(y)
\end{equation} 

In appendix \ref{App1} we perform this multiplication, obtaining

\begin{equation}
G(y)=1+\sum_{m=1}^{\infty}H_{m}^{(N)}\frac{y^{m}}{m!}
\end{equation}
where the coefficients $H_{m}^{(N)}$ can be expressed in terms of the symbols $\mathcal{C}_{n}^{m}$ introduced in refs \cite{Castro} and \cite{Castro3}, in particular 

\begin{equation}\label{HN}
H_{m}^{(N)}=\sum_{n=1}^{m}\mathcal{C}_{n}^{m}\left[\frac{\mathcal{N}_{n}^{(N)}}{N!}-\mathcal{D}_{n}\right]
\end{equation}

The symbols $\mathcal{C}_{n}^{m}$ in terms of the composition notations used in this paper are

\begin{align}\label{Cnm}
\mathcal{C}_{m}^{m}&=1 \\ \mathcal{C}_{n}^{m}&=\frac{m!}{n!}\sum_{i=1}^{m-n}(-1)^{i}\sum_{c \in \mathfrak{C}_{i}(m-n)}F(c)\,\,\,\,\,\, \mathrm{if} \,\,\,\,\, n<m \,\,\,(n \in \mathbb{N})
\end{align}

Therefore, the bivariate generating function $Z(x,y)$ is

\begin{equation}
Z(x,y)=g(y)\left(1+\sum_{N=1}^{\infty}\left\{1+\sum_{m=1}^{\infty}H_{m}^{(N)}\frac{y^{m}}{m!}\right\}\frac{x^{N}}{N!}\right)
\end{equation}

\section{The bivariate connected generating function $W(x,y)$}

In QFT, It is a well-known fact that the connected Feynman diagrams can be obtained from the logarithm of the generating functional of all the Feynman diagrams. In statistical mechanics and probability theory, the logarithm of the partition function (or the moment-generating function) is the Helmholtz free energy (or the cumulant-generating function). In our case (zero-dimensional field theory) this relation is maintained, and the bivariate function

\begin{equation}
W(x,y)=\log Z(x,y),
\end{equation}
which is the generating function of the number of connected Feynman diagrams. Particularly

\begin{equation}\label{GConn}
W(x,y)=\log g(y)+\log\left[\left(1+\sum_{N=1}^{\infty}\left\{1+\sum_{m=1}^{\infty}H_{m}^{(N)}\frac{y^{m}}{m!}\right\}\frac{x^{N}}{N!}\right)\right]
\end{equation}

The generating function of connected vacuum Feynman graph is $$w(y)=\sum_{m=0}^{\infty}\mathcal{D}_{c\,m}\frac{y^{m}}{m!}=\log g(y)=\log\left(1+\sum_{m=1}^{\infty}\frac{(2m)!}{m!}y^{m}\right)$$ with $\mathcal{D}_{c\,m}$ the number of connected vacuum graph of order $m$. Using the expansion of $$\log(1+x)=\sum_{k=1}^{\infty}\frac{(-1)^{k+1}}{k}x^{k}$$ and, by simple multiplication of power series, we can obtain explicitly the numbers  $\mathcal{D}_{c\,m}$ by the formula

\begin{equation}\label{vac}
\mathcal{D}_{c\,m}=m!\sum_{i=1}^{m}\frac{(-1)^{i+1}}{i}\sum_{c \in \mathfrak{C}_{i}(m)}F(c)
\end{equation}

Here, we have $\mathcal{D}_{c\,0}=1$ by definition. We observe that \ref{vac} is the solution of the recurrence obtained by the inverse problem $g(y)=\exp w(y)$ (see appendix of \cite{Castro3}). The recurrence in question is 

\begin{equation}
\mathcal{D}_{m}=\sum_{k=1}^{m}\frac{(m-1)!}{(i-1)!(m-1)!}\mathcal{D}_{c\,i}\mathcal{D}_{m-i}
\end{equation}

The second term of (\ref{GConn}) is more subtle, we can still apply the logarithm expansion formula, the difference lies in the coefficients $H_{m}^{(N)}$ since these are arbitrary finite sums and realize the intrinsic power series multiplication, it seems somewhat challenging. Fortunately, we can express the final result in generalizations of the symbols $\mathcal{C}_{n}^{m}$ used in \cite{Castro} and \cite{Castro3}, allowing the asymptotic expansion for the number of connected diagrams with external legs $\mathcal{N}_{c\,m}^{(N)}$. The exact expression for $\mathcal{N}_{c\,m}^{(N)}$ is obtained from

\begin{align}
\sum_{N=1}^{\infty}\left(\sum_{m=N-1}^{\infty}\mathcal{N}_{c\,m}^{(N)}\frac{y^{m}}{m!}\right)\frac{x^{N}}{N!}&=\log\left[\left(1+\sum_{N=1}^{\infty}\left\{1+\sum_{m=1}^{\infty}H_{m}^{(N)}\frac{y^{m}}{m!}\right\}\frac{x^{N}}{N!}\right)\right] \nonumber \\ &= \sum_{k=1}^{\infty}\frac{(-1)^{k+1}}{k}\left[\sum_{N=1}^{\infty}\left\{1+\sum_{m=1}^{\infty}H_{m}^{(N)}\frac{y^{m}}{m!}\right\}\frac{x^{N}}{N!}\right]^{k}
\end{align}

We expand the terms in the right-hand side and associates with the corresponding terms in the left side, finding $\mathcal{N}_{c\,m}^{(N)}$. Particularly, for $N=1,2$ and 3, we have 

\begin{align}
\mathcal{N}_{c\,m}^{(1)}&=H_{m}^{(1)} \\ \mathcal{N}_{c\,m}^{(2)}&=H_{m}^{(2)}-2H_{m}^{(1)}-H_{m}^{(1,1)} \label{Nc2} \\ \mathcal{N}_{c\,m}^{(3)}&= H_{m}^{(3)}+3H_{m}^{(1)}-3H_{m}^{(2)}+6H_{m}^{(1,1)}-3H_{m}^{(1,2)}+2H_{m}^{(1,1,1)} \label{Nc3}
\end{align}

Where, in these equations, we use the following definition

\begin{equation}\label{Hn1nj}
H_{m}^{(n_{1},n_{2},\cdots,n_{j})}=\sum_{k_{1},k_{2},\cdots,k_{j}=1}^{m-j+1}\delta_{k_{1}+k_{2}+\cdots+k_{j},m}\frac{m!}{k_{1}!k_{2}!\cdots k_{j}!}H_{k_{1}}^{(n_{1})}H_{k_{2}}^{(n_{2})}\cdots H_{k_{j}}^{(n_{j})}
\end{equation}

In appendix \ref{App1} we prove the following simplified formula, first note that the factor order of the indices $(n_{1},n_{2},\cdots,n_{j})$ in (\ref{Hn1nj}) does not matter. With this in mind, we take the non-decreasing order in the indices. Therefore, the last index $n_{j}$ is the greatest value of the indices. We obtain

\begin{equation}\label{HnMod}
H_{m}^{(n_{1},n_{2},\cdots,n_{j})}=(-1)^{j+1}\sum_{n=1}^{m-j+1}\langle \mathcal{C}_{n}^{m}\rangle_{n_{1}n_{2}\cdots n_{j-1}}\left[\frac{\mathcal{N}_{n}^{(n_{j})}}{n_{j}!}-\mathcal{D}_{n}\right]
\end{equation}

with $\langle \mathcal{C}_{n}^{m}\rangle_{n_{1}n_{2}\cdots n_{k}}$ a generalization of the symbols (\ref{Cnm}). Particularly

\begin{equation}\label{CMod}
\langle \mathcal{C}_{n}^{m}\rangle_{n_{1}n_{2}\cdots n_{k}}=\frac{m!}{n!}\sum_{i=k}^{m-n}(-1)^{i}\binom{i}{k}\sum_{c \in \mathfrak{C}_{i}(m-n)}f_{n_{1}}(a_{1})f_{n_{2}}(a_{2})\cdots f_{n_{k}}(a_{k})F(c)
\end{equation}
where, the function $f_{n_{r}}(a_{r})$ is

\begin{equation}
f_{n_{r}}(a_{r})= \left[\binom{2a_{r}+n_{r}}{n_{r}}-1\right]
\end{equation} 

The first values of $f_{n_{r}}(a)$ for $n_{r}=1,2,3\cdots$ are

\begin{align}
f_{1}(a)&=2a \\
f_{2}(a)&=3a+2a^{2} \\
f_{3}(a)&=\frac{11}{3}a+4a^{2}+\frac{4}{3}a^{3} \\
&\,\,\,\vdots \nonumber
\end{align}

We calculate, explicitily $\mathcal{N}_{c\,m}^{(N)}$ for $N=1,2$ and $3$ and show the formulas for $N=4,5$. For $N=1$ we have

\begin{equation}\label{1}
\mathcal{N}_{c\,m}^{(1)}=H_{m}^{(1)}=\sum_{n=1}^{m}\mathcal{C}_{n}^{m}\left[\mathcal{N}_{n}^{(1)}-\mathcal{D}_{n}\right]
\end{equation}
which was obtained in ref\cite{Castro}. For $N=2$, from (\ref{Nc2}), we obtain $\mathcal{N}_{c\,m}^{(2)}$. As was shown in \cite{Castro3}, $\mathcal{N}_{c\,m}^{(2)}$ can be written only in terms of the symbols $\mathcal{C}_{n}^{m}$. For see this, note that $H_{m}^{(1,1)}$ is (see (\ref{Hn1nj}))

\begin{equation}
H_{m}^{(1,1)}=-\sum_{n=1}^{m-1} \langle\mathcal{C}_{n}^{m}\rangle_{1}\left[\mathcal{N}_{n}^{(1)}-\mathcal{D}_{n}\right]=-\sum_{n=1}^{m} 2(m-n)\mathcal{C}_{n}^{m}\left[\mathcal{N}_{n}^{(1)}-\mathcal{D}_{n}\right]
\end{equation}
since

\begin{align}
\langle\mathcal{C}_{n}^{m}\rangle_{1}&=\frac{m!}{n!}\sum_{i=1}^{m-n}(-1)^{i}\binom{i}{1}\sum_{c \in \mathfrak{C}_{i}(m-n)}f_{1}(a_{1})F(c)=\frac{m!}{n!}\sum_{i=1}^{m-n}(-1)^{i}\binom{i}{1}\sum_{c \in \mathfrak{C}_{i}(m-n)}f_{1}(a_{2})F(c) \nonumber \\
&=\frac{m!}{n!}\sum_{i=1}^{m-n}(-1)^{i}i\sum_{c \in \mathfrak{C}_{i}(m-n)}\left[\frac{f_{1}(a_{1})+f_{1}(a_{2})+\cdots+f_{1}(a_{i})}{i}\right]F(c)\nonumber \\ &=\frac{m!}{n!}\sum_{i=1}^{m-n}(-1)^{i}\sum_{c \in \mathfrak{C}_{i}(m-n)}\left[2a_{1}+2a_{2}+\cdots+2a_{i}\right]F(c)= 2(m-n)\frac{m!}{n!}\sum_{i=1}^{m-n}(-1)^{i}\sum_{c \in \mathfrak{C}_{i}(m-n)}F(c) \nonumber \\ &=2(m-n)\mathcal{C}_{n}^{m}.
\end{align}

Using (\ref{Nc2}) we obtain

\begin{equation}
\mathcal{N}_{c\,m}^{(2)}=\sum_{n=1}^{m}\mathcal{C}_{n}^{m}\left[\frac{\mathcal{N}_{n}^{(2)}}{2!}-\mathcal{D}_{n}\right]-2\sum_{n=1}^{m}\mathcal{C}_{n}^{m}\left[\mathcal{N}_{n}^{(1)}-\mathcal{D}_{n}\right]+\sum_{n=1}^{m}2(m-n)\mathcal{C}_{n}^{m}\left[\mathcal{N}_{n}^{(1)}-\mathcal{D}_{n}\right]
\end{equation}

It is convenient to factor the symbols $\mathcal{C}_{n}^{m}$ and express the result in terms of a dominant factor to obtain the asymptotic expansion. For this, we write

\begin{equation}
\frac{\mathcal{N}_{n}^{(2)}}{2!}-\mathcal{D}_{n}=\left(\frac{\mathcal{N}_{n}^{(2)}}{2!}-\mathcal{N}_{n}^{(1)}\right)+\left(\mathcal{N}_{n}^{(1)}-\mathcal{D}_{n}\right)
\end{equation}
as $\left(\mathcal{N}_{n}^{(1)}-\mathcal{D}_{n}\right)=2n(2n)!$ and $$\left(\frac{\mathcal{N}_{n}^{(2)}}{2!}-\mathcal{N}_{n}^{(1)}\right)=\frac{2n+1}{2}(2n)(2n)!,$$ we obtain

\begin{equation}
\left(\mathcal{N}_{n}^{(1)}-\mathcal{D}_{n}\right)=\frac{2}{2n+1}\left(\frac{\mathcal{N}_{n}^{(2)}}{2!}-\mathcal{N}_{n}^{(1)}\right)
\end{equation}

This allows to write $\mathcal{N}_{c\,m}^{(2)}$ as 

\begin{equation}\label{2}
\mathcal{N}_{c\,m}^{(2)}=\sum_{n=1}^{m}\frac{4m-2n-1}{2n+1}\mathcal{C}_{n}^{m}\left(\frac{\mathcal{N}_{n}^{(2)}}{2!}-\mathcal{N}_{n}^{(1)}\right).
\end{equation}
In accordance with ref\cite{Castro3}.

\subsection{Cases $N>2$} 

The procedure used in the case $N=2$ can be generalized for the other cases. We may write explicitly all the terms of (\ref{Nc3}) and find the respective formula. For that we use 

\begin{equation}
\frac{\mathcal{N}_{n}^{(k)}}{k!}-\mathcal{D}_{n}=\left(\frac{\mathcal{N}_{n}^{(k)}}{k!}-\frac{\mathcal{N}_{n}^{(k-1)}}{(k-1)!}\right)+\left(\frac{\mathcal{N}_{n}^{(k-1)}}{(k-1)!}-\frac{\mathcal{N}_{n}^{(k-2)}}{(k-2)!}\right)+\cdots+\left(\mathcal{N}_{n}^{(1)}-\mathcal{D}_{n}\right)
\end{equation}
and

\begin{equation}
\left(\frac{\mathcal{N}_{n}^{(k)}}{k!}-\frac{\mathcal{N}_{n}^{(k-1)}}{(k-1)!}\right)=(2n)!(2n)\frac{(2n+1)}{2}\frac{(2n+2)}{3}\cdots\frac{(2n+k-1)}{k}
\end{equation}

The terms in question are

\begin{align}
H_{m}^{(3)}=\sum_{n=1}^{m}\frac{4n^{2}+12n+11}{(2n+2)(2n+1)}\mathcal{C}_{n}^{m}\left(\frac{\mathcal{N}_{n}^{(3)}}{3!}-\frac{\mathcal{N}_{n}^{(2)}}{2!}\right);\,\,\,\, 3H_{m}^{(1)}=\sum_{n=1}^{m}\frac{18}{(2n+2)(2n+1)}\mathcal{C}_{n}^{m}\left(\frac{\mathcal{N}_{n}^{(3)}}{3!}-\frac{\mathcal{N}_{n}^{(2)}}{2!}\right) \nonumber \\ -3H_{m}^{(2)}=-\sum_{n=1}^{m}\frac{18n+27}{(2n+2)(2n+1)}\mathcal{C}_{n}^{m}\left(\frac{\mathcal{N}_{n}^{(3)}}{3!}-\frac{\mathcal{N}_{n}^{(2)}}{2!}\right);\,\,\,\, 6H_{m}^{(1,1)}=-\sum_{n=1}^{m-1}\frac{36}{(2n+2)(2n+1)}\langle\mathcal{C}_{n}^{m}\rangle_{1}\left(\frac{\mathcal{N}_{n}^{(3)}}{3!}-\frac{\mathcal{N}_{n}^{(2)}}{2!}\right) \nonumber \\ -3H_{m}^{(2,1)}=\sum_{n=1}^{m-1}\frac{18n+27}{(2n+2)(2n+1)}\langle\mathcal{C}_{n}^{m}\rangle_{1}\left(\frac{\mathcal{N}_{n}^{(3)}}{3!}-\frac{\mathcal{N}_{n}^{(2)}}{2!}\right);\,\,\,\, 2H_{m}^{(1,1,1)}=\sum_{n=1}^{m-2}\frac{12}{(2n+2)(2n+1)}\langle\mathcal{C}_{n}^{m}\rangle_{11}\left(\frac{\mathcal{N}_{n}^{(3)}}{3!}-\frac{\mathcal{N}_{n}^{(2)}}{2!}\right)
\end{align}

Using $\langle\mathcal{C}_{n}^{m}\rangle_{1}=2(m-n)\mathcal{C}_{n}^{m}$, we obtain for (\ref{Nc3})

\begin{align}\label{3}
\mathcal{N}_{c\,m}^{(3)}=&\sum_{n=1}^{m}\frac{(n-1)(2n-1)}{(n+1)(2n+1)}\mathcal{C}_{n}^{m}\left(\frac{\mathcal{N}_{n}^{(3)}}{3!}-\frac{\mathcal{N}_{n}^{(2)}}{2!}\right)+\sum_{n=1}^{m}\frac{9(m-n)(2n-1)}{(n+1)(2n+1)}\mathcal{C}_{n}^{m}\left(\frac{\mathcal{N}_{n}^{(3)}}{3!}-\frac{\mathcal{N}_{n}^{(2)}}{2!}\right)\nonumber \\ &+\sum_{n=1}^{m-2}\frac{6}{(n+1)(2n+1)}\langle\mathcal{C}_{n}^{m}\rangle_{11}\left(\frac{\mathcal{N}_{n}^{(3)}}{3!}-\frac{\mathcal{N}_{n}^{(2)}}{2!}\right)
\end{align}
\subsubsection{Case $N=4$}

For $N=4$ we have

\begin{align}
\mathcal{N}_{c\,m}^{(4)}=&H_{m}^{(4)}-4H_{m}^{(3,1)}-4H_{m}^{(3)}-3H_{m}^{(2,2)}+12H_{m}^{(2,1,1)}+24H_{m}^{(2,1)}+6H_{m}^{(2)}\nonumber\\ &-6H_{m}^{(1,1,1,1)}-24H_{m}^{(1,1,1)}-24H_{m}^{(1,1)}-4H_{m}^{(1)} 
\end{align}

Repeating the procedure, an exact calculation gives

\begin{align}\label{4}
\mathcal{N}_{c\,m}^{(4)}=&\sum_{n=1}^{m}\frac{(2n-1)(2n-3)(n-1)}{(2n+1)(2n+3)(n+1)}\mathcal{C}_{n}^{m}\left(\frac{\mathcal{N}_{n}^{(4)}}{4!}-\frac{\mathcal{N}_{n}^{(3)}}{3!}\right)+\sum_{n=1}^{m}\frac{16(m-n)(4n^{2}-24n-7)}{(2n+1)(2n+3)(n+1)}\mathcal{C}_{n}^{m}\left(\frac{\mathcal{N}_{n}^{(4)}}{4!}-\frac{\mathcal{N}_{n}^{(3)}}{3!}\right) \nonumber \\ & +\sum_{n=1}^{m-1}\frac{18}{(2n+1)(n+1)}\langle\mathcal{C}_{n}^{m}\rangle_{2}\left(\frac{\mathcal{N}_{n}^{(4)}}{4!}-\frac{\mathcal{N}_{n}^{(3)}}{3!}\right)+\sum_{n=1}^{m-2}\frac{72(2n-1)}{(2n+3)(2n+1)(n+1)}\langle\mathcal{C}_{n}^{m}\rangle_{11}\left(\frac{\mathcal{N}_{n}^{(4)}}{4!}-\frac{\mathcal{N}_{n}^{(3)}}{3!}\right)\nonumber \\ &+ \sum_{n=1}^{m-3}\frac{72}{(2n+3)(2n+1)(n+1)}\langle\mathcal{C}_{n}^{m}\rangle_{111}\left(\frac{\mathcal{N}_{n}^{(4)}}{4!}-\frac{\mathcal{N}_{n}^{(3)}}{3!}\right)
\end{align}

The exact formula for $N=5$ is

\begin{align}\label{5}
\mathcal{N}_{c\,m}^{(5)}=&\sum_{n=1}^{m}\frac{(2n-1)(2n-3)(n-1)(n-2)}{(2n+1)(2n+3)(n+1)(n+2)}\mathcal{C}_{n}^{m}\left(\frac{\mathcal{N}_{n}^{(5)}}{5!}-\frac{\mathcal{N}_{n}^{(4)}}{4!}\right)\nonumber\\&+\sum_{n=1}^{m}\frac{25(m-n)(4n^{3}-44n^{2}+131n+89)}{(2n+1)(2n+3)(n+1)(n+2)}\mathcal{C}_{n}^{m}\left(\frac{\mathcal{N}_{n}^{(5)}}{5!}-\frac{\mathcal{N}_{n}^{(4)}}{4!}\right)\nonumber \\&+\sum_{n=1}^{m-1}\frac{100(2n^{2}-3n-8)}{(2n+1)(2n+3)(n+1)(n+2)}\langle\mathcal{C}_{n}^{m}\rangle_{2}\left(\frac{\mathcal{N}_{n}^{(5)}}{5!}-\frac{\mathcal{N}_{n}^{(4)}}{4!}\right)\nonumber \\&+\sum_{n=1}^{m-2}\frac{200(2n^{2}-21n-8)}{(2n+1)(2n+3)(n+1)(n+2)}\langle\mathcal{C}_{n}^{m}\rangle_{11}\left(\frac{\mathcal{N}_{n}^{(5)}}{5!}-\frac{\mathcal{N}_{n}^{(4)}}{4!}\right)\nonumber \\&+\sum_{n=1}^{m-2}\frac{450}{(2n+1)(n+1)(n+2)}\langle\mathcal{C}_{n}^{m}\rangle_{21}\left(\frac{\mathcal{N}_{n}^{(5)}}{5!}-\frac{\mathcal{N}_{n}^{(4)}}{4!}\right)\nonumber \\&+\sum_{n=1}^{m-3}\frac{900(2n-1)}{(2n+1)(2n+3)(n+1)(n+2)}\langle\mathcal{C}_{n}^{m}\rangle_{111}\left(\frac{\mathcal{N}_{n}^{(5)}}{5!}-\frac{\mathcal{N}_{n}^{(4)}}{4!}\right)\nonumber \\&+\sum_{n=1}^{m-4}\frac{720}{(2n+1)(2n+3)(n+1)(n+2)}\langle\mathcal{C}_{n}^{m}\rangle_{1111}\left(\frac{\mathcal{N}_{n}^{(5)}}{5!}-\frac{\mathcal{N}_{n}^{(4)}}{4!}\right)
\end{align}

The other cases can be easily calculated (using a computer). However, more and more terms appear and the expressions are quite cumbersome for the polynomials in $n$. In particular, for arbitrary $N$, we have $p(N)$ terms, with $p(N)$ the number of different partitions of $N$.

\section{Asymptotic expansions}

The exact equations (\ref{1}), (\ref{2}), (\ref{3}), (\ref{4}) and (\ref{5}) may pose some difficuties when used to calculate the number of Feynman diagram if $m$ is large. However, as was shown in Ref.\cite{Castro3}, these expressions allow to find many terms of the corresponding asymptotic expansions. Here, following Ref.\cite{Castro3} we implement computationally the methods used in this reference to calculate many asymptotics terms. The strategy is simple: 

\begin{itemize}
\item Find the term in the exact formula that gives the principal contribution.
\item Factore this term with respect to the other exact formula terms.
\item Expand in Taylor series the factorized non-principal contribution.
\end{itemize}

As was shown in \cite{Castro3}, for $m\to \infty$, the number of terms in the exact formula tends to infinity. However, only a finite number of them contribute to the $a$-th asymptotic term with $a \in \mathbb{N}$ {\it finite}. Also, we calculate some asymptotic multinomial centered contribution (see Ref.\cite{Castro3}) in each case.

\subsection{$N=0$, Vacuum diagrams}

From formula (\ref{vac}) is evident that the dominant term correspond to $i=1$, being $(2m)!$ the principal contribution. Factorizing them, we obtain

\begin{equation}\label{vac2}
\mathcal{D}_{c\,m}=(2m)!\left[1+\frac{m!}{(2m)!}\sum_{i=2}^{m}\frac{(-1)^{i+1}}{i}\sum_{c \in \mathfrak{C}_{i}(m)}F(c)\right]
\end{equation}

The sum in the square bracket term is over the compositions $$\{m-1,1\};\{1,m-1\};\{m-2,2\};\{2,m-2\};\{m-2,1,1\};\{1,m-2,1\};\cdots$$

The principal asymptotic expansion is obtained by calculating the Taylor series in $m=\infty$ for each composition-term and adding the respective series. The point is that for each asymptotic expansion term only a finite number of composition terms contribute. We implement computationally this calculus in Appendix \ref{App2}. The principal asymptotic expansion is

\begin{equation}
\mathcal{D}_{c\,m}=(2m)!\left[1-\frac{1}{2m}-\frac{3}{4m^{2}}-\frac{19}{8m^{3}}-\frac{191}{16m^{4}}-\frac{2551}{32m^{5}}-\frac{41935}{64m^{6}}-\cdots\right]
\end{equation}

As in ref \cite{Castro3}, we can calculate other compositions contributions of $m \to \infty$ called in ref \cite{Castro3} multinomial centered contributions.  For example, the binomial centered contribution is given by the following compositions of $m$

\begin{equation}\label{Comp}
\{\frac{m}{2},\frac{m}{2}\};\{\frac{m}{2}-1,\frac{m}{2},1\};\{\frac{m}{2},\frac{m}{2}-1,1\};\cdots
\end{equation}

Introducing these compositions in the formula (\ref{vac2}) and taking the taylor expansion in $m \to \infty$ we obtain the binomial centered contibution

\begin{equation}\label{Binom}
-\frac{(2m)!}{2^{m}\sqrt{2}}\left[1-\frac{33}{8m}-\frac{1599}{128m^{2}}-\frac{98683}{1024m^{3}}-\frac{37584917}{32768m^{4}}-\frac{4542533551}{262144 m^5}- \frac{1306413864411}{4194304m^6}-\cdots\right]
\end{equation}

In reference \cite{Castro3} for each $n$-nomial contribution we introduce a different asymptotic expansion depending on the nature of the positive integer $m$ as $m-n$ ($\mathrm{mod}$  $n$). For simplicity, we only consider the expansion $m\equiv n$ ($\mathrm{mod}$ $n$) for all the $n$-nomial centered contributions ((In formula (\ref{Binom}) we have $n=2$)). The next $n$-nomial contributions are, for $n=3$

\begin{equation}
\frac{(2m)!2}{3^{m+1}}\left[1-\frac{83}{6 m}-\frac{2023}{36 m^2}-\frac{540553}{648 m^3}-\frac{70424971}{3888 m^4}-\frac{10981940885}{23328 m^5}-\frac{5898719946727}{419904 m^6}-\cdots\right]
\end{equation}
and for $n=4$

\begin{equation}
-\frac{(2m)!}{4^{m}\sqrt{2}}\left[1-\frac{261}{8m} - \frac{15847}{128m^2} - \frac{3780271}{
 1024m^3} - \frac{4349723845}{32768m^4} - \frac{1384519231451}{
 262144m^5} - \frac{977871757263603}{4194304m^6}-\cdots\right].
\end{equation}
In particular, exist infinite $n$-nomial centered asymptotics contributions. We resume all this asymptotics contributions in the next table

\begin{table}[H]
	\centering
	\begin{tabular}{| c | c |c |c |c |c |c |c |}
		\hline
		& Principal & $1/m$ & $1/m^{2}$ & $1/m^{3}$ & $1/m^{4}$ & $1/m^{5}$ & $1/m^{6}$   \\ \hline $n=1$ & $(2m)!$ & $\frac{1}{2}$ & $\frac{3}{4}$ & $\frac{19}{8}$ & $\frac{191}{16}$ & $\frac{2551}{32}$ & $\frac{41935}{64}$  \\ \hline $n=2$ & $-\frac{(2m)!}{2^{m}\sqrt{2}}$ & $\frac{33}{8}$ & $\frac{1599}{128}$ & $\frac{98683}{1024}$ & $\frac{37584917}{32768}$ & $\frac{4542533551}{262144}$ & $\frac{1306413864411}{4194304}$ \\ \hline $n=3$ & $\frac{(2m)!2}{3^{m+1}}$ & $\frac{83}{6}$ & $\frac{2023}{36}$ & $\frac{540553}{648}$ & $\frac{70424971}{3888}$ & $\frac{10981940885}{23328}$ & $\frac{5898719946727}{419904}$\\ \hline  $n=4$ & $-\frac{(2m)!}{4^{m}\sqrt{2}}$ & $\frac{261}{8}$ & $\frac{15847}{128}$ & $\frac{3780271}{
 1024}$ & $\frac{4349723845}{32768}$ & $\frac{1384519231451}{
 262144}$ & $\frac{977871757263603}{4194304}$\\ \hline
		\end{tabular}
	\caption{\label{Table1}Respective coefficients in the asymptotic expansion for $N=0$ .}
\end{table}

\subsection{$N \neq 0$, Diagrams with external legs}

The same strategy can be implemented with formulas (\ref{1}), (\ref{2}), (\ref{3}), (\ref{4}) and (\ref{5}). In all these formulas the principal contribution is given by the term in the first sum at $n=m$. Note that in this term we have $\mathcal{C}_{m}^{m}=1$. In ref \cite{Castro3} we have obtain the asymptotic expansion for $\mathfrak{h}_{m}^{(1)}$ and $\mathfrak{h}_{m}^{(2)}$ (see expression (\ref{h})). In ref \cite{Castro3} we write the corresponding factor $(2m)!/m!$ in $\mathfrak{h}_{m}^{(N)}$ as

$$\frac{(2m)!}{m!}=m!\frac{(2m)!}{m!m!}=m!\binom{2m}{m}$$
and there is used an asymptotic expansion for $\binom{2m}{m}$, which must be multiplied by the asymptotic expansions obtained here.

Another important remark, in ref \cite{Castro3} the multinomial centered asymptotics expansion ($n \geq 2$) are incompleted since some terms of the exact formula were ignored. Here we consider all the neglected terms. We show our result in the next tables:    

\begin{table}[H]
	\centering
	\begin{tabular}{| c | c |c |c |c |c |c |c |}
		\hline
		& Principal & $1/m$ & $1/m^{2}$ & $1/m^{3}$ & $1/m^{4}$ & $1/m^{5}$ & $1/m^{6}$   \\ \hline $n=1$ & $(2m)!(2m)$ & $\frac{1}{2}$ & $\frac{3}{4}$ & $\frac{19}{8}$ & $\frac{191}{16}$ & $\frac{2551}{32}$ & $\frac{41935}{64}$  \\ \hline $n=2$ & $-\frac{(2m)!(2m)}{2^{m}\sqrt{2}}$ & $\frac{33}{8}$ & $\frac{1599}{128}$ & $\frac{98683}{1024}$ & $\frac{37584917}{32768}$ & $\frac{4542533551}{262144}$ & $\frac{1306413864411}{4194304}$ \\ \hline $n=3$ & $\frac{(2m)!(4m)}{3^{m+1}}$ & $\frac{83}{6}$ & $\frac{2023}{36}$ & $\frac{540553}{648}$ & $\frac{70424971}{3888}$ & $\frac{10981940885}{23328}$ & $\frac{5898719946727}{419904}$\\ \hline  $n=4$ & $-\frac{(2m)!(2m)}{4^{m}\sqrt{2}}$ & $\frac{261}{8}$ & $\frac{15847}{128}$ & $\frac{3780271}{
 1024}$ & $\frac{4349723845}{32768}$ & $\frac{1384519231451}{
 262144}$ & $\frac{977871757263603}{4194304}$ \\ \hline
		\end{tabular}
	\caption{\label{Table2}Respective coefficients in the asymptotic expansion for $N=1$ .}
\end{table}
for example, the table \ref{Table2} express the following asymptotics for $\mathcal{N}_{c\,m}^{(1)}$

\begin{align}
\mathcal{N}_{c\,m}^{(1)}=&(2m)!(2m)\left[1-\frac{1}{2m}-\cdots-\frac{41935}{64m^6}-\cdots\right]-\frac{(2m)!(2m)}{2^{m}\sqrt{2}}\left[1-\frac{33}{8m}-\cdots\right] \nonumber \\ &+\frac{(2m)!(4m)}{3^{m+1}}\left[1-\frac{83}{6m}-\cdots\right]-\frac{(2m)!(2m)}{4^{m}\sqrt{2}}\left[1-\frac{261}{8m}-\cdots\right]+\cdots
\end{align}
for the others $N$ we have

\begin{table}[H]
	\centering
	\begin{tabular}{| c | c |c |c |c |c |c |c |}
		\hline
		& Principal & $1/m$ & $1/m^{2}$ & $1/m^{3}$ & $1/m^{4}$ & $1/m^{5}$ & $1/m^{6}$   \\ \hline $n=1$ & $(2m)!m(2m-1)$ & $\frac{1}{2}$ & $\frac{7}{4}$ & $\frac{35}{8}$ & $\frac{315}{16}$ & $\frac{4063}{32}$ & $\frac{65875}{64}$  \\ \hline $n=2$ & $-\frac{(2m)!(3m)(2m-1)}{2^{m+1}\sqrt{2}}$ & $\frac{95}{24}$ & $\frac{6053}{384}$ & $\frac{335117}{3072}$ & $\frac{41864677}{32768}$ & $\frac{15124159777}{786432}$ & $\frac{4342579796905}{12582912}$ \\ \hline $n=3$ & $\frac{(2m)!(10m)(2m-1)}{3^{m+2}}$ & $\frac{409}{30}$ & $\frac{11567}{180}$ & $\frac{2825147}{3240}$ & $\frac{367944611}{19440}$ & $\frac{57423636091}{116640}$ & $\frac{30856162470899}{2099520}$\\ \hline  $n=4$ & $-\frac{(2m)!(7m)(2m-1)}{4^{m+1}\sqrt{2}}$ & $\frac{1815}{56}$ & $\frac{125289}{896}$ & $\frac{26865149}{7168}$ & $\frac{31135765875}{229376}$ & $\frac{9929674629241}{
 1835008}$ & $\frac{7019179014307085}{29360128}$ \\ \hline
		\end{tabular}
	\caption{\label{Table3}Respective coefficients in the asymptotic expansion for $N=2$ .}
\end{table}  

\begin{table}[H]
	\centering
	\begin{tabular}{| c | c |c |c |c |c |c |c |}
		\hline
		& Principal & $1/m$ & $1/m^{2}$ & $1/m^{3}$ & $1/m^{4}$ & $1/m^{5}$ & $1/m^{6}$   \\ \hline $n=1$ & $(2m)!\frac{m}{3}(2m-1)(2m-2)$ & $\frac{1}{2}$ & $\frac{15}{4}$ & $\frac{67}{8}$ & $\frac{515}{16}$ & $\frac{6511}{32}$ & $\frac{106891}{64}$  \\ \hline $n=2$ & $-(2m)!\frac{5m(2m-1)(2m-2)}{2^{m+1}\times3\sqrt{2}}$ & $\frac{153}{40}$ & $\frac{2967}{128}$ & $\frac{730427}{5120}$ & $\frac{258987193}{163840}$ & $\frac{6200256955}{262144}$ & $\frac{8899805020047}{20971520}$ \\ \hline $n=3$ & $(2m)!\frac{62m(2m-1)(2m-2)}{3^{m+4}}$ & $\frac{2483}{186}$ & $\frac{90181}{1116}$ & $\frac{19324681}{20088}$ & $\frac{2489331889}{120528}$ & $\frac{389719812857}{723168}$ & $\frac{209788775838769}{13017024}$\\ \hline  $n=4$ & $-(2m)!\frac{m(2m-1)(2m-2)}{4^{m-1}\times3\sqrt{2}}$ & $\frac{513}{16}$ & $\frac{2747}{16}$ & $\frac{7947125}{2048}$ & $\frac{4647075355}{32768}$ & $\frac{2979220269043}{524288
}$ & $\frac{527637111526281}{2097152}$ \\ \hline
		\end{tabular}
	\caption{\label{Table4}Respective coefficients in the asymptotic expansion for $N=3$ .}
\end{table}  

\begin{table}[H]
	\centering
	\begin{tabular}{| c | c |c |c |c |c |c |c |}
		\hline
		& Principal & $1/m$ & $1/m^{2}$ & $1/m^{3}$ & $1/m^{4}$ & $1/m^{5}$ & $1/m^{6}$   \\ \hline $n=1$ & $(2m)!\frac{m(2m-1)(2m-2)(2m-3)}{12}$ & $\frac{1}{2}$ & $\frac{27}{4}$ & $\frac{127}{8}$ & $\frac{863}{16}$ & $\frac{10627}{32}$ & $\frac{177343}{64}$  \\ \hline $n=2$ & $-(2m)!\frac{35m(2m-1)(2m-2)(2m-3)}{2^{m+5}\times3\sqrt{2}}$ & $\frac{207}{56}$ & $\frac{154797}{4480}$ & $\frac{1010879}{5120}$ & $\frac{457473011}{229376}$ & $\frac{273390683141}{9175040}$ & $\frac{79181536650081}{146800640}$ \\ \hline $n=3$ & $(2m)!\frac{71m(2m-1)(2m-2)(2m-3)}{3^{m+4}\times2}$ & $\frac{5569}{426}$ & $\frac{271757}{2556}$ & $\frac{51508175}{46008}$ & $\frac{6421619153}{276048}$ & $\frac{1012202986195}{1656288}$ & $\frac{547529808276557}{29813184}$\\ \hline  $n=4$ & $-(2m)!\frac{679m(2m-1)(2m-2)(2m-3)}{4^{m+4}\times3\sqrt{2}}$ & $\frac{171963}{5432}$ & $\frac{19060513}{86912}$ & $\frac{2861770561}{695296}$ & $\frac{3355407464611}{22249472}$ & $\frac{1085932334589125}{177995776}$ & $\frac{772486899922478229}{2847932416}$ \\ \hline
		\end{tabular}
	\caption{\label{Table5}Respective coefficients in the asymptotic expansion for $N=4$ .}
\end{table}

\begin{table}[H]
	\centering
	\begin{tabular}{| c | c |c |c |c |c |c |c |}
		\hline
		& Principal & $1/m$ & $1/m^{2}$ & $1/m^{3}$ & $1/m^{4}$ & $1/m^{5}$ & $1/m^{6}$   \\ \hline $n=1$ & $(2m)!\frac{m(2m-1)(2m-2)(2m-3)(2m-4)}{60}$ & $\frac{1}{2}$ & $\frac{43}{4}$ & $\frac{239}{8}$ & $\frac{1551}{16}$ & $\frac{17491}{32}$ & $\frac{289135}{64}$  \\ \hline $n=2$ & $-(2m)!\frac{21m(2m-1)(2m-2)(2m-3)(2m-4)}{2^{m+5}\times5\sqrt{2}}$ & $\frac{257}{72}$ & $\frac{402257}{8064}$ & $\frac{17977229}{64512}$ & $\frac{1716536537}{688128}$ & $\frac{617217081673}{16515072}$ & $\frac{182510071595413}{264241152}$ \\ \hline $n=3$ & $(2m)!\frac{517m(2m-1)(2m-2)(2m-3)(2m-4)}{3^{m+5}\times10}$ & $\frac{39731}{3102}$ & $\frac{2620231}{18612}$ & $\frac{453118381}{335016}$ & $\frac{53173092307}{2010096}$ & $\frac{8481478160825}{12060576}$ & $\frac{4635195165206599}{217090368}$\\ \hline  $n=4$ & $-(2m)!\frac{1969m(2m-1)(2m-2)(2m-3)(2m-4)}{4^{m+4}\times15\sqrt{2}}$ & $\frac{492189}{15752}$ & $\frac{71257863}{252032}$ & $\frac{9032609159}{2016256}$ & $\frac{10457264010645}{64520192}$ & $\frac{3436211431133539}{516161536}$ & $\frac{2461472084613042227}{8258584576}$ \\ \hline
		\end{tabular}
	\caption{\label{Table6}Respective coefficients in the asymptotic expansion for $N=5$ .}
\end{table} 

In appendix \ref{App2} we explicitly show as we calculate all this coefficients after sixth perturbation order.  

\section{Conclusion and discussion}

In this work, we have implemented a pure formal series calculation scheme in order to obtain exact formulas for the number of connected Wick contractions with an arbitrary number of external legs. Also, these exact formulas make possible the calculus of many asymptotics terms of the corresponding asymptotics expansion. The standard techniques for the enumeration of Wick contractions are in connection with the resolution of formal integrals in the complex plane. In particular, the classical reference \cite{Cvitanovic} uses, for a class of quantum field theories, their respective generating functionals, changing the associated functional integral to conventional integrals and solving them. Their numerical solution corresponds to the number of Wick Contractions. The asymptotics is obtained there using the steepest descent method. Another approach (see \cite{Molinari}, \cite{Pavlyukh}) is the use of Dyson and many-body relations which can be solved numerically, or transforming them into ordinary differential equations, suitable for obtaining the asymptotic expansions. Here we used another equivalent approach, by means of a pure generating function approach (without involved formal integrals) derived in [17], from the Wick theorem. That procedure allowed us to obtain exact formulas for an arbitrary number of external legs, which has the advantage of showing explicitly the intricacies present in the limit $m\to \infty$. The exact formula is a finite alternating sum whose terms are analytics in $m=\infty$. In the limit $m\to \infty$ the number of term in the exact formula tends to infinite and the asymptotic expansion when $m \to \infty$ have convergence radius equal to zero.

About the combinatorial side of this work, we have applied standard generating
function calculus to obtain exact non-closed expressions (finite sums), whose terms are indexed by the set of numerical compositions. The form of these exact formulas has similarities with combinatorial formulas obtained by sieve methods (see chapter 2 of Ref.\cite{Stanley}) which carry to the exact result by a ``finite set" of successive approximations (in our case, indexed by numerical compositions). The ``sieve formulas" need $2^{m-1}$ successive approximations in order to achieve the exact result. Considering all these composition terms in the limit $m \to \infty$, we obtain an infinite alternating sum of different asymptotic expansions; The dominant asymptotic expansion corresponds to the usual one given in the literature. We show explicitly the presence of these other asymptotic series contributions negligible respect to the principal asymptotic series. This can be interpreted explicitly as a non-perturbative contribution deductible directly from our exact calculus. Our calculus is possible through the total knowledge of the generating function $Z(x,y)$ in (\ref{Z}). Applying the $Log\left(Z(x,y)\right)$ we obtain the particular dependence respect to the compositions of $m$. Particularly,  compositions with two or more elements depending on $m$ (see the composition in (\ref{Comp})) are the source of the ``non-perturbative" negligible contributions. Here the quotes come to the fact that this is a toy model for QFT which preserves some of the characteristics of non-zero-dimensional quantum field theory. In non-zero-dimensional QFT the non-perturbative effects come from instanton contribution in the Feynman integral functional approach. Would-be interessing if our methods (or part of them) can be generalized to non-zero dimensional QFT. We hope that this work can be generalized in this direction.    

\section*{ACKNOWLEDGMENTS}

E. R. Castro is grateful to the CBPF institution for the infrastructure offered during the writing process of this work.

I. Roditi is grateful for the hospitality of the Quantum Information Group at
ETHZ, where this work has been completed.

This study was partially financed by the Coordena\c{c}\~ao de
Aperfei\c{c}oamento de Pessoal de N\'ivel Superior-Brasil (CAPES)-Finance
Code 001.

\bibliography{RefsArticle}

\begin{thebibliography}{18}
\expandafter\ifx\csname natexlab\endcsname\relax\def\natexlab#1{#1}\fi
\expandafter\ifx\csname bibnamefont\endcsname\relax
  \def\bibnamefont#1{#1}\fi
\expandafter\ifx\csname bibfnamefont\endcsname\relax
  \def\bibfnamefont#1{#1}\fi
\expandafter\ifx\csname citenamefont\endcsname\relax
  \def\citenamefont#1{#1}\fi
\expandafter\ifx\csname url\endcsname\relax
  \def\url#1{\texttt{#1}}\fi
\expandafter\ifx\csname urlprefix\endcsname\relax\def\urlprefix{URL }\fi
\providecommand{\bibinfo}[2]{#2}
\providecommand{\eprint}[2][]{\url{#2}}

\bibitem[{\citenamefont{Flajolet and Sedgewick}(2009)}]{Flajolet}
\bibinfo{author}{\bibfnamefont{P.}~\bibnamefont{Flajolet}} \bibnamefont{and}
  \bibinfo{author}{\bibfnamefont{R.}~\bibnamefont{Sedgewick}},
  \emph{\bibinfo{title}{Analytic Combinatorics}} (\bibinfo{publisher}{Cambridge
  university press}, \bibinfo{year}{2009}), ISBN \bibinfo{isbn}{9780521898065}.

\bibitem[{\citenamefont{Baxter}(1982)}]{Baxter}
\bibinfo{author}{\bibfnamefont{R.}~\bibnamefont{Baxter}},
  \emph{\bibinfo{title}{Exactly Solved Models in Statistical Mechanics}}
  (\bibinfo{publisher}{ACADEMIC PRESS LIMITED}, \bibinfo{year}{1982}), ISBN
  \bibinfo{isbn}{0-12-083180-5}.

\bibitem[{\citenamefont{Cvitanovic et~al.}(1978)\citenamefont{Cvitanovic,
  Lautrup, and Pearson}}]{Cvitanovic}
\bibinfo{author}{\bibfnamefont{P.}~\bibnamefont{Cvitanovic}},
  \bibinfo{author}{\bibfnamefont{B.}~\bibnamefont{Lautrup}}, \bibnamefont{and}
  \bibinfo{author}{\bibfnamefont{R.~B.} \bibnamefont{Pearson}},
  \bibinfo{journal}{Phys. Rev. D} \textbf{\bibinfo{volume}{18}},
  \bibinfo{pages}{1939} (\bibinfo{year}{1978}).

\bibitem[{\citenamefont{Argyres et~al.}(2001)\citenamefont{Argyres, van
  Hameren, Kleiss, and Papadopoulos}}]{Argyres}
\bibinfo{author}{\bibfnamefont{E.}~\bibnamefont{Argyres}},
  \bibinfo{author}{\bibfnamefont{A.}~\bibnamefont{van Hameren}},
  \bibinfo{author}{\bibfnamefont{R.}~\bibnamefont{Kleiss}}, \bibnamefont{and}
  \bibinfo{author}{\bibfnamefont{C.}~\bibnamefont{Papadopoulos}},
  \bibinfo{journal}{The European Physical Journal C - Particles and Fields}
  \textbf{\bibinfo{volume}{19}}, \bibinfo{pages}{567 } (\bibinfo{year}{2001}).

\bibitem[{\citenamefont{Molinari}(2005)}]{Molinari}
\bibinfo{author}{\bibfnamefont{L.~G.} \bibnamefont{Molinari}},
  \bibinfo{journal}{Phys. Rev. B} \textbf{\bibinfo{volume}{71}},
  \bibinfo{pages}{113102} (\bibinfo{year}{2005}).

\bibitem[{\citenamefont{Pavlyukh and Hubner}(2007)}]{Pavlyukh}
\bibinfo{author}{\bibfnamefont{Y.}~\bibnamefont{Pavlyukh}} \bibnamefont{and}
  \bibinfo{author}{\bibfnamefont{W.}~\bibnamefont{Hubner}},
  \bibinfo{journal}{Journal of Mathematical Physics}
  \textbf{\bibinfo{volume}{48}}, \bibinfo{pages}{052109}
  (\bibinfo{year}{2007}).

\bibitem[{\citenamefont{Borinsky}(2017)}]{Borinsky}
\bibinfo{author}{\bibfnamefont{M.}~\bibnamefont{Borinsky}},
  \bibinfo{journal}{Annals of Physics} \textbf{\bibinfo{volume}{385}},
  \bibinfo{pages}{95} (\bibinfo{year}{2017}).

\bibitem[{\citenamefont{Abdesselam}(2002)}]{Abdesselam}
\bibinfo{author}{\bibfnamefont{A.}~\bibnamefont{Abdesselam}},
  \emph{\bibinfo{title}{Feynman diagrams in algebraic combinatorics}}
  (\bibinfo{year}{2002}), \eprint{arXiv:math/0212121}.

\bibitem[{\citenamefont{Jackson et~al.}(2017)\citenamefont{Jackson, Kempf, and
  Morales}}]{Jackson}
\bibinfo{author}{\bibfnamefont{D.}~\bibnamefont{Jackson}},
  \bibinfo{author}{\bibfnamefont{A.}~\bibnamefont{Kempf}}, \bibnamefont{and}
  \bibinfo{author}{\bibfnamefont{A.~H.} \bibnamefont{Morales}},
  \bibinfo{journal}{J. Phys. A: Math. Theor} \textbf{\bibinfo{volume}{50}},
  \bibinfo{pages}{225201} (\bibinfo{year}{2017}).

\bibitem[{\citenamefont{Yeats}(2017)}]{Yeats}
\bibinfo{author}{\bibfnamefont{K.}~\bibnamefont{Yeats}},
  \emph{\bibinfo{title}{A Combinatorial Perspective on Quantum Field Theory}},
  Springer Briefs in Mathematical Physics (\bibinfo{publisher}{Springer},
  \bibinfo{year}{2017}).

\bibitem[{\citenamefont{Geloun and Ramgoolam}(2014)}]{Geloun}
\bibinfo{author}{\bibfnamefont{J.~B.} \bibnamefont{Geloun}} \bibnamefont{and}
  \bibinfo{author}{\bibfnamefont{S.}~\bibnamefont{Ramgoolam}},
  \bibinfo{journal}{Annales Del institut Henri Poincare D}
  \textbf{\bibinfo{volume}{1}}, \bibinfo{pages}{77} (\bibinfo{year}{2014}).

\bibitem[{\citenamefont{Prunotto et~al.}(2018)\citenamefont{Prunotto, Alberico,
  and Czerski}}]{Prunotto}
\bibinfo{author}{\bibfnamefont{A.}~\bibnamefont{Prunotto}},
  \bibinfo{author}{\bibfnamefont{W.}~\bibnamefont{Alberico}}, \bibnamefont{and}
  \bibinfo{author}{\bibfnamefont{P.}~\bibnamefont{Czerski}},
  \bibinfo{journal}{Open Physics} \textbf{\bibinfo{volume}{16}}
  (\bibinfo{year}{2018}).

\bibitem[{\citenamefont{Krishna et~al.}(2018)\citenamefont{Krishna, Labelle,
  and Shramchenko}}]{Gopala}
\bibinfo{author}{\bibfnamefont{K.~G.} \bibnamefont{Krishna}},
  \bibinfo{author}{\bibfnamefont{P.}~\bibnamefont{Labelle}}, \bibnamefont{and}
  \bibinfo{author}{\bibfnamefont{V.}~\bibnamefont{Shramchenko}},
  \bibinfo{journal}{Nuclear Physics B} \textbf{\bibinfo{volume}{936}},
  \bibinfo{pages}{668} (\bibinfo{year}{2018}).

\bibitem[{\citenamefont{Castro}(2018)}]{Castro}
\bibinfo{author}{\bibfnamefont{E.}~\bibnamefont{Castro}},
  \bibinfo{journal}{Journal of Mathematical Physics}
  \textbf{\bibinfo{volume}{59}}, \bibinfo{pages}{023503}
  (\bibinfo{year}{2018}).

\bibitem[{\citenamefont{Walsh and Lehman}(1972)}]{Walsh}
\bibinfo{author}{\bibfnamefont{T.}~\bibnamefont{Walsh}} \bibnamefont{and}
  \bibinfo{author}{\bibfnamefont{A.}~\bibnamefont{Lehman}},
  \bibinfo{journal}{Journal of Combinatorial Theory, Series B}
  \textbf{\bibinfo{volume}{13}}, \bibinfo{pages}{192} (\bibinfo{year}{1972}).

\bibitem[{\citenamefont{Arqu\`es and B\'eraud}(2000)}]{Arques}
\bibinfo{author}{\bibfnamefont{D.}~\bibnamefont{Arqu\`es}} \bibnamefont{and}
  \bibinfo{author}{\bibfnamefont{J.~F.} \bibnamefont{B\'eraud}},
  \bibinfo{journal}{Discrete Mathematics, Elsevier}
  \textbf{\bibinfo{volume}{200}}, \bibinfo{pages}{1} (\bibinfo{year}{2000}).

\bibitem[{\citenamefont{Castro and Roditi}(2019)}]{Castro3}
\bibinfo{author}{\bibfnamefont{E.~R.} \bibnamefont{Castro}} \bibnamefont{and}
  \bibinfo{author}{\bibfnamefont{I.}~\bibnamefont{Roditi}},
  \bibinfo{journal}{Journal of Physics A: Mathematical and Theoretical}
  \textbf{\bibinfo{volume}{52}}, \bibinfo{pages}{5401} (\bibinfo{year}{2019}).

\bibitem[{\citenamefont{Stanley}(2012)}]{Stanley}
\bibinfo{author}{\bibfnamefont{R.~P.} \bibnamefont{Stanley}},
  \emph{\bibinfo{title}{Enumerative combinatorics, Vol I}}
  (\bibinfo{publisher}{Cambridge University Press}, \bibinfo{year}{2012}).

\end{thebibliography}

\appendix

\section{Proofs of some results used in the paper}\label{App1}

In this appendix, we proceed to prove some formulas established in the paper

\subsection{Proof of equation (\ref{hm})}

The equation is valid for $m=1$, suppose that it is valid until a párticular $m$. By formula (\ref{Dm}), we have

\begin{equation}
\sum_{n=0}^{m+1}\frac{\mathcal{D}_{n}}{n!}h_{m-n+1}=0
\end{equation}
or

\begin{equation}\label{hinduc}
h_{m+1}\mathcal{D}_{0}=h_{m+1}=-\sum_{n=0}^{m}\frac{(2n)!}{n!}h_{m-n+1}=-\sum_{n=0}^{m}\frac{(2n)!}{n!}\sum_{i=1}^{m-n+1}(-1)^{i}\sum_{c \in \mathfrak{C}_{i}(m-n+1)}F(c)
\end{equation}

In the last step, we use the induction hypothesis. The multiplication of $-(2n)!/n$ by $h_{m-n+1}$ generates terms of the desired formula since they have the same form and are represented by some compositions of $m+1$. Particularly, all the compositions of $m+1$ can be obtained as follows: Fix the first element to $n=1$ and take all the compositions of $m$ for the other elements. This gives all the compositions of $m+1$ with the first element equal to 1. Repeat the procedure for $n=2,3,\cdots,m+1$, this exhausts all the possibilities. It is clear that the formula (\ref{hinduc}) implements this procedure. Therefore 

\begin{equation}
h_{m+1}=\sum_{i=1}^{m+1}(-1)^{i}\sum_{c \in \mathfrak{C}_{i}(m+1)}F(c)
\end{equation} 
which proves (\ref{hm})

\subsection{Proof of equation (\ref{HN})}

From formula (\ref{ghy}), we have for $m>0$

\begin{align}
H_{m}^{(N)}&=m!\sum_{n=0}^{m}\frac{\mathcal{N}_{n}^{(N)}}{N!n!}h_{m-n}=m!\left[h_{m}+\left(\sum_{n=1}^{m}\frac{(2n+N)!}{N!n!}h_{m-n}\right)\right] \nonumber \\ &=m!\left[h_{m}+\left(\sum_{n=0}^{m-1}\frac{(2(m-n)+N)!}{N!(m-n)!}h_{n}\right)\right]
\end{align}

A second of reflexion is sufficient to perceive that, for $N=0$, the term between parentheses is

\begin{equation}
\left(\sum_{n=0}^{m-1}\frac{(2(m-n)+0)!}{0!(m-n)!}h_{n}\right)=-h_{m}
\end{equation}

This is not the case for other values of $N$. However, for  $N\neq 0$, the formula has the same form as $h_{m}$ with the following substitution for the dependence in an element of the composition (say the last) in the formula $F(c)$ 

\begin{equation}
\frac{(2a_{i})!}{a_{i}!}\to \frac{(2a_{i}+N)!}{a_{i}!N!}
\end{equation}

Therefore

\begin{align}
H_{m}^{(N)}&=m!\sum_{i=1}^{m}(-1)^{i}\sum_{c \in \mathfrak{C}_{i}(m)}\left(\prod_{j=1}^{i-1}\frac{(2a_{j})!}{a_{j}!}\right)\left[\frac{(2a_{i})!}{a_{i}!}-\frac{(2a_{i}+N)!}{a_{i}!N!}\right] \nonumber \\ &=m!\sum_{i=1}^{m}(-1)^{i-1}\sum_{c \in \mathfrak{C}_{i}(m)}\left(\prod_{j=1}^{i-1}\frac{(2a_{j})!}{a_{j}!}\right)\frac{1}{a_{i}!}\left[\frac{\mathcal{N}_{a_{i}}^{(N)}}{N!}-\mathcal{D}_{a_{i}}\right] \nonumber \\
\end{align}

Remember that $c=(a_{1},a_{2},\cdots,a_{i})$ and $a_{1}+a_{2}+\cdots+a_{i}=m$. If we substitute $a_{i}\to n$, we can write 

\begin{equation}
H_{m}^{(N)}=\sum_{n=1}^{m}\left\{\frac{m!}{n!}\sum_{i=1}^{m-n}(-1)^{i}\sum_{c \in \mathfrak{C}_{i}(m-n)}F(c)\right\}\left[\frac{\mathcal{N}_{n}^{(N)}}{N!}-\mathcal{D}_{n}\right]
\end{equation}
and using equation (\ref{Cnm}), we obtain

\begin{equation}
H_{m}^{(N)}=\sum_{n=1}^{m}\mathcal{C}_{n}^{m}\left[\frac{\mathcal{N}_{n}^{(N)}}{N!}-\mathcal{D}_{n}\right]
\end{equation}

\subsection{Proof of equations (\ref{HnMod}) and (\ref{CMod})}

From formulas (\ref{Hn1nj}) and (\ref{HN}) we can write

\begin{equation}\label{A10}
H_{m}^{(n_{1},n_{2},\cdots,n_{j})}=\sum_{k_{1},k_{2},\cdots,k_{j}=1}^{m-j+1}\delta_{k_{1}+k_{2}+\cdots+k_{j},m}m!\frac{H_{k_{1}}^{(n_{1})}}{k_{1}!}\frac{H_{k_{2}}^{(n_{2})}}{k_{2!}}\cdots \frac{H_{k_{j}}^{(n_{j})}}{k_{j}!}
\end{equation}
where

\begin{equation}
\frac{H_{k_{s}}^{(n_{r})}}{k_{s}!}=\left[\frac{\mathcal{N}_{k_{s}}^{(n_{r})}}{n_{i}!k_{s}!}-\frac{\mathcal{D}_{k_{s}}}{k_{s}!}\right]+\sum_{n=1}^{k_{s}-1}\left\{\sum_{i=1}^{k_{s}-n}(-1)^{i}\sum_{c \in \mathfrak{C}_{i}(k_{s}-n)}F(c)\right\}\left[\frac{\mathcal{N}_{n}^{(n_{r})}}{n_{r}!n!}-\frac{\mathcal{D}_{n}}{n!}\right]
\end{equation}

For $r=j$ we leave this formula in this form, but if $r<j$ we write

\begin{equation}
\frac{H_{k_{s}}^{(n_{r})}}{k_{s}!}=\left[f_{n_{r}}(k_{s})\frac{(2k_{s})!}{k_{s}!}\right]+\sum_{n=1}^{k_{s}-1}\left\{\sum_{i=1}^{k_{s}-n}(-1)^{i}\sum_{c \in \mathfrak{C}_{i}(k_{s}-n)}F(c)\right\}\left[f_{n_{r}}(n)\frac{(2n)!}{n!}\right]
\end{equation}
with

\begin{equation}
f_{n_{r}}(a_{r})= \left[\binom{2a_{r}+n_{r}}{n_{r}}-1\right].
\end{equation}

The terms of $H_{k_{s}}^{(n_{r})}/k_{s}!$ are indexed by the compositions of $k_{s}$. Realizing the multiplication in (\ref{A10}) we obtain a set of terms all of them in the form

\begin{equation}\label{A14}
(-1)^{j+1}\frac{m!}{n!}(-1)^{i}f_{n_{1}}(a_{1})f_{n_{2}}(a_{2})\cdots f_{n_{j-1}}(a_{j-1})F(c)\left[\frac{\mathcal{N}_{n}^{(n_{j})}}{n_{j}!}-\mathcal{D}_{n}\right]
\end{equation}
with $c=(a_{1},a_{2},\cdots,a_{j-1},\cdots,a_{i})$, $i\geq j-1$ and the composition $c \in \mathfrak{C}_{i}(m-n)$. The factor $(-1)^{j+1}$ is necessary to guarantee that with each element of the composition $c$ be associated a multiplicative factor $(-1)$. It is evident that each term of (\ref{A10}) is associated with a composition of $m$. Also, gives an arbitrary composition of $m$ with $i\geq j$ and $i+1$ elements, always we will find this represented by some term in (\ref{A10}). For see this we write

\begin{equation}\label{A15}
m=a_{1}+a_{2}+\cdots+a_{i}+n=[a_{1}+\cdots+a_{i_{1}}]_{1}+[a_{i_{1}+1}+\cdots+a_{i_{2}}]_{2}+\cdots+[a_{i_{j-1}+1}+\cdots+a_{i}+n]_{j}
\end{equation}
where the indexed brackets $[\cdots]_{s}$ contain a composition of $k_{s}$ which represent some term of $H_{k_{s}}^{(n_{s})}/k_{s}!$ in the product (\ref{A10}). If $i>j-1$ the way to group the composition (\ref{A15}) in $j$ brackets is not unique. In particular, there are $\binom{i}{j-1}$ ways to group the $i+1$ elements of (\ref{A15}) in $j$ brackets. To prove this consider $i=j-1$, we have

\begin{equation}
m=a_{1}+a_{2}+\cdots+a_{i}+n=[a_{1}]_{1}+[a_{2}]_{2}+\cdots+[a_{j-1}]_{j-1}+[n]_{j}
\end{equation}
which is the unique possible grouping (particularly, we have $\binom{j-1}{j-1}=1$). Suppose that the claim is valid until an arbitrary $i$. For $i+1$,  we can group in $j$ brackets with the new element $a_{i+1}$ in two ways:

\begin{itemize}
\item Add $a_{i+1}$ on the last bracket of the compositions of $m-a_{i+1}$ which is grouped in $j$ brackets and interchange with $n$ 
\begin{align}
m&=[a_{1}+\cdots+a_{i_{1}}]_{1}+[a_{i_{1}+1}+\cdots+a_{i_{2}}]_{2}+\cdots+[a_{i_{j-1}+1}+\cdots+a_{i}+n+a_{i+1}]_{j}\nonumber \\&=[a_{1}+\cdots+a_{i_{1}}]_{1}+[a_{i_{1}+1}+\cdots+a_{i_{2}}]_{2}+\cdots+[a_{i_{j-1}+1}+\cdots+a_{i}+a_{i+1}+n]_{j}
\end{align}
\item Create a new bracket containing only the element $a_{i+1}$, add this to the compositions of $m-a_{i+1}$ which is grouped in $j-1$ brackets and interchange with $n$

\begin{align}
m&=[a_{1}+\cdots+a_{i_{1}}]_{1}+[a_{i_{1}+1}+\cdots+a_{i_{2}}]_{2}+\cdots+[a_{i_{j-1}+1}+\cdots+a_{i}+n]_{j-1}+[a_{i+1}]_{j}\nonumber \\&=[a_{1}+\cdots+a_{i_{1}}]_{1}+[a_{i_{1}+1}+\cdots+a_{i_{2}}]_{2}+\cdots+[a_{i_{j-1}+1}+\cdots+a_{i}+a_{i+1}]_{j-1}+[n]_{j}
\end{align}
\end{itemize}

Using the induction hypothesis,  the number of different ways to group in $j$ brackets the composition of $m$ with $i+2$ elements is simply  

\begin{equation}
\binom{i}{j-1}+\binom{i}{j-2}=\binom{i+1}{j-1}
\end{equation} 

Which proves our claim of possible groupings. Therefore, adding over all the possibles compositions of $m$ and over the possible groupings in $j$ brackets, We obtain from (\ref{A14})

\begin{equation}
(-1)^{j+1}\sum_{n=1}^{m-j+1}\frac{m!}{n!}\sum_{i=j-1}^{m-n}(-1)^{i}\binom{i}{j-1}\sum_{c \in \mathfrak{C}_{i}(m-n)}f_{n_{1}}(a_{1})f_{n_{2}}(a_{2})\cdots f_{n_{j-1}}(a_{j-1})F(c)\left[\frac{\mathcal{N}_{n}^{(n_{j})}}{n_{j}!}-\mathcal{D}_{n}\right]=H_{m}^{(n_{1},n_{2},\cdots,n_{j})}
\end{equation}
Which proves (\ref{HnMod}) and (\ref{CMod}).

\section{Computational implementation for the asymptotic expansion calculus}\label{App2}

In this appendix, we implement computationally the calculus of the asymptotics coefficients until the sixth perturbation order. For this, we use the program MATHEMATICA for calculating all the compositions and the corresponding terms that contributes to the asymptotic expansion until sixth order. 

First, we calculate the compositions of $m$ used for obtaining the principal asymptotic contribution ($n=1$) in all the exacts formulas.  

{\color{blue}\begin{lstlisting}
    	T[m] = DeleteCases[Flatten[Table[
    	Table[Permutations[Flatten[{m - a + 1, 
    	IntegerPartitions[a - 1][[k]]}]], 
      	{k, 1, Length[IntegerPartitions[a - 1]]}], {a, 1, 7}], 2], {m}]
    	\end{lstlisting}}
    	
 The compositions of $m$ that contribute to the binomial centered asymptotic expansion ($n=2$) are contained in {\color{blue} \begin{lstlisting}
    	T2[m]
    	\end{lstlisting}} and are explicitly calculated in this way:
    	
{\color{blue}\begin{lstlisting}
    	P2[a_] :=Join[Permutations[Flatten[{IntegerPartitions[a, {1}], 0}]], 
  	Flatten[Table[Permutations[IntegerPartitions[a, {2}][[k]]], {k, 1, 
     	Length[IntegerPartitions[a, {2}]]}], 1]];
    	\end{lstlisting}}    	 	 

{\color{blue}\begin{lstlisting}
    	P2[0] = {{0, 0}};
    	\end{lstlisting}}    	 

{\color{blue}\begin{lstlisting}
    	F2[m_, a_] := Table[{m/2 - P2[a][[k, 1]], m/2 - P2[a][[k, 2]]}, {k, 1, 
   	Length[P2[a]]}];
    	\end{lstlisting}}    	 
and finally

{\color{blue}\begin{lstlisting}
    	T2[m] = DeleteDuplicates[Flatten[Table[
    	Table[Permutations[
      	Join[F2[m, a - 1][[k]], IntegerPartitions[a - 1][[l]]]], {k, 1, 
      	Length[F2[m, a - 1]]}, {l, 1, 
      	Length[IntegerPartitions[a - 1]]}], {a, 1, 7}], 3]]
    	\end{lstlisting}}    

Similarly, the other compositions corresponding to the other asymptotics centered contributions (particularly $n=3$ and $n=4$) are calculated by 

{\color{blue}\begin{lstlisting}
    	P3[a_] := Join[Permutations[Flatten[{IntegerPartitions[a, {1}], 0, 0}]], 
  	Flatten[Table[Permutations[Join[IntegerPartitions[a, {2}][[k]],
  	{0}]], {k, 1,Length[IntegerPartitions[a, {2}]]}], 1], 
  	Flatten[Table[Permutations[IntegerPartitions[a, {3}][[k]]], {k, 1, 
     	Length[IntegerPartitions[a, {3}]]}], 1]];
    	\end{lstlisting}}    	 	 

{\color{blue}\begin{lstlisting}
    	P3[0] = {{0, 0, 0}};
    	\end{lstlisting}}    	 

{\color{blue}\begin{lstlisting}
    	F3[m_, a_] := Table[{Factor[m/3 - P3[a][[k, 1]]],
    	Factor[m/3 - P3[a][[k, 2]]], Factor[m/3 - P3[a][[k, 3]]]},
    	{k, 1, Length[P3[a]]}];
    	\end{lstlisting}}    	 

{\color{blue}\begin{lstlisting}
    	T3[m] = DeleteDuplicates[
  		Flatten[Table[Table[Permutations[
        Join[F3[m, a - 1][[k]], IntegerPartitions[a - 1][[l]]]], {k, 1, 
      	Length[F3[m, a - 1]]}, {l, 1, 
      	Length[IntegerPartitions[a - 1]]}], {a, 1, 7}], 3]]
    	\end{lstlisting}}
    	
and    

{\color{blue}\begin{lstlisting}
    	P4[a_] := Join[Permutations[Flatten[{IntegerPartitions[a, {1}],
    	0, 0, 0}]],Flatten[Table[
    	Permutations[Join[IntegerPartitions[a, {2}][[k]],
    	{0, 0}]], {k, 1, Length[IntegerPartitions[a, {2}]]}], 1], 
  	Flatten[Table[Permutations[Join[IntegerPartitions[a, {3}][[k]],
  	{0}]], {k, 1, Length[IntegerPartitions[a, {3}]]}], 1], 
  	Flatten[Table[Permutations[IntegerPartitions[a, {4}][[k]]],
  	{k, 1, Length[IntegerPartitions[a, {4}]]}], 1]];
    	\end{lstlisting}}    	 	 

{\color{blue}\begin{lstlisting}
    	P4[0] = {{0, 0, 0, 0}};
    	\end{lstlisting}}    	 

{\color{blue}\begin{lstlisting}
    	F4[m_, a_] := Table[{Factor[m/4 - P4[a][[k, 1]]],
    	Factor[m/4 - P4[a][[k, 2]]], Factor[m/4 - P4[a][[k, 3]]],
    	Factor[m/4 - P4[a][[k, 4]]]}, {k, 1, 
   	Length[P4[a]]}];
    	\end{lstlisting}}    	 

{\color{blue}\begin{lstlisting}
    	T4[m] = DeleteDuplicates[
  	Flatten[Table[Table[Permutations[Join[F4[m, a - 1][[k]], 
  	IntegerPartitions[a - 1][[l]]]], {k, 1, 
      	Length[F4[m, a - 1]]}, {l, 1, 
      	Length[IntegerPartitions[a - 1]]}], {a, 1, 7}], 3]]
    	\end{lstlisting}} 	    

Each list of these compositions must be introduced in the respective exact formulas. For example, in formula (\ref{3}) for $\mathcal{N}_{c\,m}^{(3)}$ denoting the lists 
{\color{blue}\begin{lstlisting}
    	T[m],T2[m],T3[m],T4[m]
 \end{lstlisting}}
generally as
{\color{blue}\begin{lstlisting}
    	Ta[m];
    	\end{lstlisting}}
    	
we have for $\mathcal{N}_{c\,m}^{(3)}$

{\color{blue}\begin{lstlisting}
    	Nc[m_] := (2 m)!*m/3 (2 m - 2) (2 m - 1) (1 +
    	Sum[((-1)^(Length[Ta[m][[i]]] + 1)*Ta[m][[i, 1]])/m*
 	m!/(2 m)!*((2 Ta[m][[i, 1]] - 1) (2 Ta[m][[i, 1]] - 2))/((2 m - 
    	1) (2 m - 2))Product[
   	(2 Ta[m][[i, r]])!/(Ta[m][[i, r]])!,{r,			    
   	1,Length[Ta[m][[i]]]}]+(((-1)^(Length[Ta[m][[i]]]+ 1)
   	*Ta[m][[i, 1]])/m)*m!/(2 m)!*(
 	18 (2 Ta[m][[i, 1]] - 1) (m - Ta[m][[i, 1]]))/((2 m - 1) (2 m - 2))
 	Product[(2 Ta[m][[i, r]])!/(Ta[m][[i, r]])!,{r,			    
   	1,Length[Ta[m][[i]]]}]+S[m, i],{i,1,Length[Ta[m]]}])
    	\end{lstlisting}}  

with
{\color{blue}\begin{lstlisting}
    	S[m_, i_] := 
 	If[Length[Ta[m][[i]]] > 
   	2, ((-1)^(Length[Ta[m][[i]]] + 1)*Ta[m][[i, 1]])/m*m!/(2 m)!*(
   	12*4*Ta[m][[i, 2]]*Ta[m][[i, 3]])/((2 m - 1) (2 m - 2))*
   	Binomial[Length[Ta[m][[i]]] - 1, 2]Product[
   	(2 Ta[m][[i, r]])!/(Ta[m][[i, r]])!,{r,			    
   	1,Length[Ta[m][[i]]]}],0]
	\end{lstlisting}}        	 

For the principal asymptotic contribution, we obtain an expression similar to this

\begin{equation}
(2m)!\frac{m}{3}(2m-2)(2m-1)\left(1+F_{1}[m]\right)
\end{equation}

For the other asymptotics centered contributions, we don't consider the factor 1 in the last parentheses. This is

\begin{equation}
(2m)!\frac{m}{3}(2m-2)(2m-1)F_{a}[m]
\end{equation}

In any case, we expand in Taylor series $F_{a}[m]$ for $m\to \infty$ using

{\color{blue}\begin{lstlisting}
    	Simplify[Series[Expand[F[m]],{m,Infinity,6}]]
    	\end{lstlisting}} 

obtaining the respective asymptotic expansion. The same procedure is valid with the others formulas.

\end{document}